\documentclass[english,aps,manuscript]{revtex4-1}
\usepackage[T1]{fontenc}
\usepackage[latin9]{inputenc}
\usepackage{float}
\usepackage{amsmath}
\usepackage{amssymb}
\usepackage{graphicx}

\makeatletter

\newcommand{\noun}[1]{\textsc{#1}}

\makeatother

\usepackage{babel}
\begin{document}

\title{Black-hole lasing in Bose-Einstein condensates: analysis of the role
of the dynamical instabilities in a nonstationary setup}

\author{J.M. Gomez Llorente and J. Plata}

\address{Departamento de F\'{\i}sica, Universidad de La Laguna,\\
 La Laguna E38204, Tenerife, Spain.}
\begin{abstract}
We present a theoretical study on the origin of some findings of recent
experiments on sonic analogs of gravitational black holes. We focus
on the realization of a black-hole lasing configuration, where the
conclusive identification of stimulated Hawking radiation requires
dealing with the implications of the nonstationary character of the
setup. To isolate the basic mechanisms responsible for the observed
behavior, we use a toy model where nonstationarity can be described
in terms of departures from adiabaticity. Our approach allows studying
which aspects of the characterization of black-hole lasing in static
models are still present in a dynamical scenario. In particular, variations
in the role of the dynamical instabilities can be traced. Arguments
to conjecture the twofold origin of the detected amplification of
sound are given: the differential effect of the instabilities on the
mean field and on the quantum fluctuations gives some clues to separate
a deterministic component from self-amplified Hawking radiation. The
role of classical noise, present in the experimental setup, is also
tackled: we discuss the emergence of differences with the effect of
quantum fluctuations when various unstable modes are relevant to the
dynamics. 
\end{abstract}
\maketitle

\section{Introduction}

The predictions on Hawking radiation (HR), i.e., on spontaneous emission
of thermal radiation from a black hole (BH)\cite{key-Hawking1,key-Hawking2},
have not had observational verification. The low temperatures of emission
constitute the main handicap to the observation. Yet, the fundamental
character of the involved physics allows searching for alternative
strategies to test the theory via the detection of similar effects
in parallel systems \cite{key-Unruh}. Specially promising are the
proposals for building sonic analogs of gravitational BHs with atomic
Bose-Einstein condensates (BECs) \cite{key-Garay1,key-Garay2,key-Barcelo,key-Gardiner}.
The basis of the proposals is the creation of a sonic BH horizon,
where the character of a flowing condensate changes from subsonic
to supersonic. As in the gravitational context, the quantum treatment
of the field, in contrast with the classical description, predicts
the emission of \emph{radiation} from the supersonic region \cite{key-Carusotto2,key-Carusotto3,key-MacherParentani}.
Given the low temperatures and the access to measure in any region
in the condensates, the proposed schemes can be expected to have practical
applicability. Indeed, a first realization was reported in \cite{key-Steinhauer3}.
Advances in this line have been achieved recently with the implementation
of a black-hole lasing (BHL) configuration \cite{key-Steinhauer1}.
In this variation of the basic scheme, a condensate flow is made to
cross twice the speed of sound. Its characteristics in a stationary
regime have been analyzed in previous theoretical work. The presence
of the second (white-hole) (WH) horizon has been predicted to lead
to amplification of spontaneous HR through a resonance mechanism similar
to that of a lasing cavity \cite{key-Corley}. The analyses are based
on the use of the Bogoliubov\textendash de Gennes (BdG) approach to
describe the excitations \cite{key-Michel1,key-Carusotto4,key-Finazzi1}.
The appearance of complex eigenvalues of the BdG operator marks the
emergence of lasing: the instabilities, associated with the imaginary
parts of the eigenfrequencies, induce the amplification of the perturbations,
in particular, the self-amplification of the quantum fluctuations.
As the amplification is frequency dependent, the thermal character
of the spontaneous HR is lost in BHL. In the experiments \cite{key-Steinhauer1},
the BH horizon was generated by displacing a step-like potential along
an atomic BEC initially at rest. In this form, the fluid was accelerated
from a subsonic to a supersonic velocity in the reference frame of
the step. The subsequent deceleration, due to the effect of the trap,
led to the second (WH) horizon. The observed magnitudes were the density
and the nonlocal density correlation function (DCF) in the supersonic
region. The evolution and spatial dependence of these observables
have been compared with classical-field simulations and predictions
of studies on static models. Interestingly, despite the nonstationary
character of the implementation, the static picture seems to provide
valuable insight into the underlying mechanisms. For instance, the
concept of dynamical instability has been used to interpret the measured
increase in the DCF. From those analyses, there is broad agreement
on the detection of lasing. Still, there is some debate on whether
the amplified radiation is actually rooted in quantum noise. (Alternative
interpretations tracing the origin to classical noise or to a deterministic
seed have been formulated from numerical simulations \cite{key-Carusotto1,key-Clark}).
A related question which requires a detailed explanation is the growth
of the mean density, absent in the predictions for a static model.
(In the analysis of this issue, the role of the nonlinear backreaction
of the excited modes on the condensate has been considered \cite{key-Michel2}).
Also, the spectral structure of the DCF must be clarified. Here, we
aim at establishing a link between the (static) theoretical approaches
and the (dynamical) experimental realization. As the questions that
have been the subject of debate have fundamental character, we have
opted for tackling them in a simple scenario, where the basic physical
mechanisms can be isolated. In particular, we deal with a toy model
where the corrections to an adiabatic approximation can already give
the clues to the effects of nonstationarity, and, in turn, to tracing
the origin of some of the observed features. The predictions of this
simplified description will be tested through a numerical simulation
of the experiment using a more complete model.

The outline of the paper is as follows. In Sec. II, we present our
model system. A nonadiabatic approach with a generic set of variable
parameters is developed in Sec. III. In Sec. IV, we evaluate the effects
of nonadiabaticity on the density and on the DCF. Differential effects
of the instabilities on quantum fluctuations and diverse types of
classical noise are tackled in Sec. V.  In Sec. VI, we go beyond the
postadiabatic scenario: numerical results are presented for the simulation
of the experimental setup with a model with no restrictions on the
time-variation regime. Finally, some general conclusions are summarized
in Sec. VI. (To give a self-contained presentation, we summarize some
results of previous descriptions of BHL in static systems in Appendix
A. Those results are taken as starting point in our study).

\section{The model system}

We consider an atomic BEC in a confining potential $V_{ex}(\mathbf{r},t)$.
We assume that the field operator of the condensate $\hat{\Psi}$
can be separated as 

\begin{equation}
\hat{\Psi}=\Psi_{0}+\hat{\delta\Psi},\label{eq:BasicAnsatz}
\end{equation}
 where $\Psi_{0}$ is a classical field and $\hat{\delta\Psi}$ is
a perturbative quantum contribution. The mean-field description of
the classical component is given by the time-dependent Gross-Pitaevskii
(GP) equation \cite{key-Stringari}

\begin{equation}
i\hbar\frac{\partial\Psi_{0}(\mathbf{r},t)}{\partial t}=\left[-\frac{\hbar^{2}}{2m}\nabla_{\mathbf{r}}^{2}+V_{ex}(\mathbf{r},t)+g\left|\Psi_{0}(\mathbf{r},t)\right|^{2}\right]\Psi_{0}(\mathbf{r},t),\label{eq:TimeDepGP}
\end{equation}
where $m$ is the mass of a condensate atom, and $g$ is the strength
that characterizes the atom-atom interaction. It is assumed that the
evolution of the quantum term $\hat{\delta\Psi}$ can be analyzed
via a linearization procedure. 

In the experimental realization, $V_{ex}(\mathbf{r},t)$ was conveniently
varied in order to implement the black-hole laser configuration, i.e.,
to achieve two crossing points between the flow velocity and the speed
of sound. Specifically, a step potential was displaced along the condensate
in the harmonic trap. The consequent acceleration (in the frame of
the step) of the atomic flow led to the emergence of the BH horizon.
Subsequently, as the trapping effect set in, the flow was decelerated
and the inner (WH) horizon was formed. Appropriate for implementing
an adiabatic approximation is to formally incorporate the time variation
in the external potential through a set of parameters $\boldsymbol{\Lambda}(t)$.
Therefore, we rewrite the confining term as $V_{ex}(\mathbf{r},\boldsymbol{\Lambda}(t))$.
Additionally, a dimensional reduction, allowed by the characteristics
of the harmonic trap, is applied: the strong confinement in the directions
transversal to the step displacement can be incorporated through effective
parameters in the equation for the reduced dynamics in the longitudinal
$x$-direction. {[}The effective monodimensional version of the trapping
potential will be denoted as $V_{ex}(x,\boldsymbol{\Lambda}(t))${]}. 

A crucial aspect of the practical arrangement is the nonstationary
character of the flow generated by the changing potential. Previous
work on the characterization of instabilities has been carried out
on models where stationary flows and permanent horizon-schemes are
assumed. In the present case, the whole process of horizon formation
presents nontrivial dynamical aspects, which constitute a handicap
to the interpretation of the findings. Actually, there is an open
debate on the relevance of different mechanisms to the observed behavior
\cite{key-Carusotto1,key-Clark,key-SteinhauerNova}. The discussion
has incorporated arguments developed from various numerical simulations.
Those studies, based on solving the GP equation, have included different
components of the setup. Early results for a completely classical
field were argued to reproduce the detected amplification. The appearance
and growth of undulations in the mean density were reported. Additionally,
the DCF, obtained with added classical noise, seemed to evolve in
agreement with the observations. The similar spatial patterns of the
density and of the DCF pointed to a common origin for the growth of
the mean values of both observables. A global conclusion was that
the emergence of amplification  in a purely classical context ruled
out the necessary identification of the observed radiation with self-amplified
HR \cite{key-Carusotto1}. In contrast with this picture, subsequent
numerical work, which included elements simulating quantum fluctuations,
revealed qualitative differences between the responses of the system
to quantum and classical noise \cite{key-SteinhauerNova}. Quantum
fluctuations were found to be necessary to account for salient spectral
features. Also, it was argued that, since the density undulation (the
\emph{ripple}) appeared before the WH formation and evolved steadily
with no qualitative changes, its origin must be different from that
of the DCF evolution, linked initially to spontaneous HR, and, developing
subsequently (different) characteristics associated with BHL. The
questions raised by this debate show the convenience of presenting
additional arguments to conform a conclusive interpretation of the
experiments. Here, we work with a simplified version of the setup
where the roles of different components of the dynamics can be singled
out. To deal with implications intrinsic to nonstationarity, we consider
a working regime where the adiabatic approximation gives the zero-order
description of the dynamics and the departures from adiabaticity are
taken  as corrections. Additionally, the quantum fluctuations from
the adiabatic stationary solutions  are assumed to be well described
by the BdG approach. The presence of classical fluctuations is also
considered. The experimental results, satisfactorily reproduced by
some of the numerical studies \cite{key-Carusotto1,key-Clark,key-SteinhauerNova},
seem to indicate a more regular and slower evolution of the flow once
the horizons have been formed. From this feature we extract some clues
to simplify the approach: instead of attempting to describe the whole
process of the emergence of the BHL, we will focus on the system once
the BHL regime has been reached. Moreover, as it will be specified
further on, in order to avoid the breakdown of the adiabatic approximation
\cite{key-Anglin}, we will concentrate on a temporal range where
no changes in the unstable or stable character of the eigenstates
of the BdG operator take place. Although the evolution in the practical
setup cannot be assumed  to correspond to a perturbed adiabatic regime,
the study of nonadiabaticity as a correction in our model will be
shown to uncover some basic mechanisms potentially relevant to general
time variations in the BHL configuration. Moreover, even though our
approach cannot account for the potential transition of the emitted
radiation from spontaneous to self-amplified HR, it can trace the
differences with the characteristics of the density growth.

\section{Non-adiabatic approach to a non-stationary black-hole lasing configuration}

Following the above considerations, we use for the field operator
in Eq. \ref{eq:BasicAnsatz} the ansatz

\begin{eqnarray}
\hat{\Psi}(x,t) & = & \Psi_{0}^{ad}(x,\boldsymbol{\Lambda}(t))+\hat{\delta\Psi}(x,t)\nonumber \\
 & = & \left[\Phi_{0}^{ad}(x,\boldsymbol{\Lambda}(t))+\hat{\delta\Phi}(x,t)\right]\exp\left[-\frac{i}{\hbar}\int_{0}^{t}\mu(\boldsymbol{\Lambda}(t^{\prime}))dt^{\prime}\right]\label{eq:AdiabaAnsatz}
\end{eqnarray}
where $\Phi_{0}^{ad}(x,\boldsymbol{\Lambda}(t))$ is the adiabatic
wavefunction, i.e., the solution to the time-independent Gross-Pitaevskii
equation for a \emph{frozen} set of parameters $\boldsymbol{\Lambda}(t)$

\begin{equation}
\left[-\frac{\hbar^{2}}{2m}\frac{\partial^{2}}{\partial x^{2}}+V_{ex}(x,\boldsymbol{\Lambda}(t))+g\left|\Phi_{0,ad}\right|^{2}-\mu(\boldsymbol{\Lambda}(t)\right]\Phi_{0}^{ad}(x,\boldsymbol{\Lambda}(t))=0.\label{eq:AdiabaGPeq}
\end{equation}
 $\mu(\boldsymbol{\Lambda}(t)$ is the related chemical potential.
Moreover, $\hat{\delta\Phi}(x,t)$ is the perturbative quantum term.
From it, we define the two-component field $\mathbf{\boldsymbol{\hat{\delta\Phi}}}$
as 

\begin{equation}
\boldsymbol{\hat{\delta\Phi}}\equiv\begin{pmatrix}\hat{\delta\Phi}\\
\hat{\delta\Phi}^{\dagger}
\end{pmatrix},\label{eq:QuanContriVector}
\end{equation}
 and, following the standard procedure to solve for it (see the Appendix),
we work initially with its classical counterpart

\begin{equation}
\boldsymbol{\delta\Phi}\equiv\begin{pmatrix}\delta\Phi\\
\delta\Phi^{*}
\end{pmatrix}.\label{eq:ClassicalQuanConV}
\end{equation}
 It is straightforwardly shown that, to first order, $\boldsymbol{\delta\Phi}$,
as does $\boldsymbol{\hat{\delta\Phi}}$, obeys the equation 

\begin{equation}
\frac{\partial\boldsymbol{\delta\Phi}}{\partial t}=-\frac{i}{\hbar}\boldsymbol{\mathcal{L}_{BdG}^{ad}}\boldsymbol{\delta\Phi}+\boldsymbol{S},\label{eq:AdBGeq}
\end{equation}
 where two contributions to the evolution can be differentiated. First,
the homogeneous part, characterized by the operator

\begin{equation}
\boldsymbol{\mathcal{L}_{BdG}^{ad}}\equiv\begin{pmatrix}H_{0}-\mu+2g\left|\Phi_{0}^{ad}\right|^{2} & g\left(\Phi_{0}^{ad}\right)^{2}\\
-g\left(\Phi_{0}^{ad*}\right)^{2} & -\left(H_{0}-\mu+2g\left|\Phi_{0}^{ad}\right|^{2}\right)
\end{pmatrix},\label{eq:AdBGOper}
\end{equation}
 where $H_{0}=-\frac{\hbar^{2}}{2m}\frac{\partial^{2}}{\partial x^{2}}+V_{ex}(x,\boldsymbol{\Lambda}(t))$,
corresponds to the evolution that would experience the perturbative
term in a strictly adiabatic regime. Note that it parallels the equation
for the perturbation in a static scenario. 

Second, the source matrix 

\begin{equation}
\boldsymbol{S}\equiv\begin{pmatrix}S\\
-S^{*}
\end{pmatrix},\label{eq:SourcVect}
\end{equation}
with  

\begin{equation}
S=-i\hbar\dot{\boldsymbol{\Lambda}}\frac{d\Phi_{0,ad}}{d\boldsymbol{\Lambda}},\label{eq:SourcFunct}
\end{equation}
 specifically incorporates departures from adiabaticity. This term
affects the dynamics irrespective of the system preparation. Actually,
it constitutes a seed for perturbations.

In the following, we will build up the general solution for the perturbative
field operator $\hat{\delta\Phi}$ as the sum of a general solution
to the homogeneous equation, which will be denoted as $\hat{\delta\Phi}^{ad}$,
and a particular solution to the complete equation, represented by
$\delta\Phi_{0}^{nad}$. Hence, we will write 
\begin{equation}
\hat{\delta\Phi}=\hat{\delta\Phi}^{ad}+\delta\Phi_{0}^{nad}.\label{eq:SumofHomandNonh}
\end{equation}

\subsection{The homogeneous equation}

The parallelism existent between the adiabatic BdG operator $\boldsymbol{\mathcal{L}_{BdG}^{ad}}$
and its counterpart $\boldsymbol{\mathcal{L}_{BdG}}$ for a strictly
static system allows applying the methods developed for the case of
a stationary flow \cite{key-Leonhardt,key-Finazzi1}, summarized in
the Appendix, to solve the homogeneous equation. Accordingly, we will
expand the field in the eigenmodes of $\boldsymbol{\mathcal{L}_{BdG}^{ad}}$.
As we are interested primarily in the effect of the instabilities
on the quantum fluctuations, we work with the quantum form of the
expansion and retain in it only the contribution of the unstable modes.
Therefore, we write 
\begin{eqnarray}
\hat{\delta\Psi}^{ad}(x,t) & = & e^{-\frac{i}{\hbar}\int_{0}^{t}\mu(\boldsymbol{\Lambda}(t^{\prime}))dt^{\prime}}\hat{\delta\Phi}^{ad}\nonumber \\
 & \sim & e^{-\frac{i}{\hbar}\int_{0}^{t}\mu(\boldsymbol{\Lambda}(t^{\prime}))dt^{\prime}}\Phi_{0}^{ad}(x,\boldsymbol{\Lambda}(t))\sum_{a}\biggl[e^{-i\vartheta_{a}(t)}\xi_{a}(x,\boldsymbol{\Lambda}(t))\hat{b}_{a}+e^{-i\vartheta_{a}^{*}(t)}\psi_{a}(x,\boldsymbol{\Lambda}(t))\hat{c}_{a}+\nonumber \\
 &  & e^{i\vartheta_{a}^{*}(t)}\eta_{a}^{*}(x,\boldsymbol{\Lambda}(t))\hat{b}_{a}^{\dagger}+e^{i\vartheta_{a}(t)}\zeta_{a}^{*}(x,\boldsymbol{\Lambda}(t))\hat{c}_{a}^{\dagger}\biggr]\label{eq:AdEXpQuan}
\end{eqnarray}
 where the operators $\hat{b}_{a}$ ($\hat{b}_{a}^{\dagger}$) and
$\hat{c}_{a}$($\hat{c}_{a}^{\dagger}$) respectively correspond to
the unstable modes

\begin{eqnarray}
\boldsymbol{V_{a}}(x,\boldsymbol{\Lambda}(t)) & = & \begin{pmatrix}\Phi_{0}^{ad}(x,\boldsymbol{\Lambda}(t))\xi_{a}(x,\boldsymbol{\Lambda}(t))\\
\Phi_{0}^{ad*}(x,\boldsymbol{\Lambda}(t))\eta_{a}(x,\boldsymbol{\Lambda}(t))
\end{pmatrix},\label{eq:ModeV}\\
\boldsymbol{Z_{a}}(x,\boldsymbol{\Lambda}(t)) & = & \begin{pmatrix}\Phi_{0}^{ad}(x,\boldsymbol{\Lambda}(t))\psi_{a}(x,\boldsymbol{\Lambda}(t))\\
\Phi_{0}^{ad*}(x,\boldsymbol{\Lambda}(t))\zeta_{a}(x,\boldsymbol{\Lambda}(t))
\end{pmatrix},\label{eq:ModeZ}
\end{eqnarray}
with respective eigenfrequencies $\lambda_{a}(\boldsymbol{\Lambda}(t))$
and $\lambda_{a}^{*}(\boldsymbol{\Lambda}(t))$ {[}$\lambda_{a}(\boldsymbol{\Lambda}(t))=\omega_{a}(\boldsymbol{\Lambda}(t))+i\Gamma_{a}(\boldsymbol{\Lambda}(t))$,
see the Appendix{]}. Moreover, we have used 

\begin{equation}
\vartheta_{a}(t)=\int_{0}^{t}\lambda_{a}(\boldsymbol{\Lambda}(t^{\prime}))dt^{\prime}.\label{eq:IntegEigenval}
\end{equation}
It is assumed that an adiabatic approximation for the wave functions
of the modes is feasible. Indeed, we can regard the nonadiabatic corrections
to the wavefunctions of the unstable modes as having a secondary effect
on the dynamics compared with the role played by the complex character
of their eigenvalues. As the applicability of this approximation can
be jeopardized by changes in the stability of the modes, we will focus
on a temporal range where no changes in the unstable or stable character
of the eigenmodes of $\boldsymbol{\mathcal{L}_{BdG}^{ad}}$ take place.
Despite its apparent restrictive character, this sound time regime
will be shown to be appropriate to uncover general characteristics
of the mechanisms underlying the observed features.

To account, in the proposed adiabatic framework, for the evolution
of a classical perturbation, e.g., of classical noise in the initial
preparation, we can use the classical version of Eq. (\ref{eq:AdEXpQuan}),
where the operators are replaced by $\mathbb{C}$-functions corresponding
to the projections of the perturbation on the unstable modes. This
analysis is postponed to  Sec. V.

\subsection{The source term}

Convenient to deal with the inhomogeneous term in Eq. (\ref{eq:AdBGeq})
is the use of the time-evolution operator $\boldsymbol{U}(t)$ associated
with $\boldsymbol{\mathcal{L}_{BdG}^{ad}}$, i.e., of the operator
defined through the equation 
\begin{equation}
\frac{d\boldsymbol{U}}{dt}=-\frac{i}{\hbar}\boldsymbol{\mathcal{L}_{BdG}^{ad}}\boldsymbol{\boldsymbol{U}}.\label{eq:UnitOper}
\end{equation}
 Using it, one can readily show that a particular solution to Eq.
(\ref{eq:AdBGeq}) is given by 
\begin{equation}
\boldsymbol{\delta\Phi_{0}^{nad}}=\boldsymbol{U}(t)\int_{0}^{t}\boldsymbol{U}^{-1}(t^{\prime})\boldsymbol{S}(t^{\prime})dt^{\prime}.\label{eq:nonAdCorr}
\end{equation}

The characterization of $\boldsymbol{U}(t)$, which, in a general
regime, can be considerably involved  because of the time dependence
of $\boldsymbol{\mathcal{L}_{BdG}^{ad}}$, is trivial in the considered
adiabatic approximation for the eigenmodes of $\boldsymbol{\mathcal{L}_{BdG}^{ad}}$.
Indeed, in the representation of (instantaneous) adiabatic eigenstates,
$\boldsymbol{U}(t)$, and, in turn, $\boldsymbol{U}^{-1}(t)$, are
diagonal. Accordingly, using Eq. (\ref{eq:IntegEigenval}), we write
the matrix elements of the operators $\boldsymbol{U}(t)$ and $\boldsymbol{U}^{-1}(t)$
in terms of the eigenvalues $\lambda_{j}(t)$ of the instantaneous
$\boldsymbol{\mathcal{L}_{BdG}^{ad}}$ as 

\begin{equation}
\boldsymbol{U}_{jk}=\exp\left(-i\int_{0}^{t}\lambda_{j}(t^{\prime})dt^{\prime}\right)\delta_{jk}=\exp\left(-i\vartheta_{j}(t)\right)\delta_{jk}\label{eq:Uelem}
\end{equation}

\begin{equation}
(\boldsymbol{U}^{-1})_{jk}=\exp\left(i\int_{0}^{t}\lambda_{j}(t^{\prime})dt^{\prime}\right)\delta_{jk}=\exp\left(i\vartheta_{j}(t)\right)\delta_{jk}.\label{eq:InverseUelement}
\end{equation}
 Consequently, the projection of Eq. (\ref{eq:nonAdCorr}) on the
eigenmodes of $\boldsymbol{\mathcal{L}_{BdG}^{ad}}$, and, in turn,
the expansion of $\delta\Phi_{0}^{nad}$ in them, is straightforward.
As we deal here with corrections to the classical field, it is the
classical version of the expansion that is applicable. Accordingly,
we write

\begin{eqnarray}
\delta\Psi_{0}^{nad}(x,t) & \sim & \exp\left(-\frac{i}{\hbar}\int_{0}^{t}\mu(\boldsymbol{\Lambda}(t^{\prime}))dt^{\prime}\right)\Phi_{0}^{ad}(x,\boldsymbol{\Lambda}(t))\times\nonumber \\
 &  & \sum_{a}\biggr[B_{a}(t)\xi_{a}(x,\boldsymbol{\Lambda}(t))+C_{a}(t)\psi_{a}(x,\boldsymbol{\Lambda}(t))+\nonumber \\
 &  & B_{a}^{*}(t)\eta_{a}^{*}(x,\boldsymbol{\Lambda}(t))+C_{a}^{*}(t)\zeta_{a}^{*}(x,\boldsymbol{\Lambda}(t))\biggl]\label{eq:NonAdexpan}
\end{eqnarray}
 where  instead of the operators $\hat{b}_{a}$ ($\hat{b}_{a}^{\dagger}$)
and $\hat{c}_{a}$ ($\hat{c}_{a}^{\dagger}$), present in the expansion
of the quantum contribution to the  field, we have here the (\emph{c-number})
time-dependent functions $B_{a}$ ($B_{a}^{*}$) and $C_{a}$ ($C_{a}^{*}$),
which give the projections of Eq. (\ref{eq:nonAdCorr}) on the eigenmodes
of $\boldsymbol{\mathcal{L}_{BdG}^{ad}}$, specifically, on the (discrete)
modes $\boldsymbol{V_{a}}$ and $\boldsymbol{Z_{a}}$. Again, as we
are interested in the effect of the instabilities, we have retained
only the contribution of the unstable (discrete) modes in the expansion.
 According to Eq. (\ref{eq:nonAdCorr}), those functions are given
by 

\begin{eqnarray}
B_{a}(t) & = & \frac{e^{-i\vartheta_{a}(t)}}{i\hbar}\int_{0}^{t}dt^{\prime}\biggl\{ e^{i\vartheta_{a}(t^{\prime})}\times\nonumber \\
 &  & \int_{-\infty}^{\infty}dx\Bigl[\Phi_{0}^{ad*}(x,\boldsymbol{\Lambda}(t))\xi_{a}^{*}(x,\boldsymbol{\Lambda}(t))S(x,t^{\prime})+\nonumber \\
 &  & \Phi_{0}^{ad}(x,\boldsymbol{\Lambda}(t))\eta_{a}^{*}(x,\boldsymbol{\Lambda}(t))S^{*}(x,t^{\prime})\Bigr]\biggr\},\label{eq:Bfunction}
\end{eqnarray}

\begin{eqnarray}
C_{a}(t) & = & \frac{e^{-i\vartheta_{a}^{*}(t)}}{i\hbar}\int_{0}^{t}dt^{\prime}\biggl\{ e^{i\vartheta_{a}^{*}(t^{\prime})}\times\nonumber \\
 &  & \int_{-\infty}^{\infty}dx\Bigl[\Phi_{0}^{ad*}(x,\boldsymbol{\Lambda}(t))\psi_{a}^{*}(x,\boldsymbol{\Lambda}(t))S(x,t^{\prime})+\nonumber \\
 &  & \Phi_{0}^{ad}(x,\boldsymbol{\Lambda}(t))\zeta_{a}^{*}(x,\boldsymbol{\Lambda}(t))S^{*}(x,t^{\prime})\Bigr]\biggr\},\label{eq:Cfunction}
\end{eqnarray}
 where the eigenfunctions $\xi_{a}(x,\boldsymbol{\Lambda}(t))$, $\eta_{a}(x,\boldsymbol{\Lambda}(t))$,
$\psi_{a}(x,\boldsymbol{\Lambda}(t))$, and $\zeta_{a}(x,\boldsymbol{\Lambda}(t))$
have the same characteristics as those corresponding to $\boldsymbol{\mathcal{L}_{BdG}}$,
introduced in the Appendix. Note that it is the projection of the
eigenmodes with the source term $\boldsymbol{S}$ that determines
the relevance  of the nonadiabatic corrections.

\section{The effects of nonadiabaticity on the density and on the nonlocal
density correlation function}

To first order in $\hat{\delta\Phi}$, the density, $\hat{\rho}=\hat{\Psi}(x,t)\hat{\Psi}^{\dagger}(x,t)$,
can be written as 

\begin{equation}
\hat{\rho}=\rho_{0}^{ad}+\hat{\rho}_{1}^{ad}+\rho_{1}^{nad},\label{eq:DensAd+NonAd}
\end{equation}
 where $\rho_{0}^{ad}(x,\boldsymbol{\Lambda}(t))=\left|\Phi_{0}^{ad}(x,\boldsymbol{\Lambda}(t))\right|^{2}$
and 

\begin{equation}
\hat{\rho}_{1}^{ad}=(\Phi_{0}^{ad}\hat{\delta\Phi}^{ad\dagger}+\textrm{h.c.}),\label{eq:den1Ad}
\end{equation}

\begin{equation}
\rho_{1}^{nad}=\left[\Phi_{0}^{ad}\delta\Phi_{0}^{nad*}+\textrm{c.c.}\right].\label{eq:dens1NonAd}
\end{equation}
One of the most conspicuous experimental features is the growth of
the density. The identification of its origin, in particular, the
characterization of its differential aspects with the mechanism responsible
for the DCF evolution, is one of the open questions. Let us see that
some clues to its understanding can be obtained by using our approach.
Specifically, we consider the system in the vacuum state $\left|0\right\rangle $
of the annihilation operators $\hat{d}_{a+}$ and $\hat{d}_{a-}$,
defined from $\hat{b}_{a}$($\hat{b}_{a}^{\dagger}$) and $\hat{c}_{a}$($\hat{c}_{a}^{\dagger}$)
as $\hat{d}_{a+}=\frac{\hat{b}_{a}+i\hat{c}_{a}}{\sqrt{2}}$, and
$\hat{d}_{a-}=\frac{\hat{b}_{a}^{\dagger}+i\hat{c}_{a}^{\dagger}}{\sqrt{2}}$.
(This is the reference vacuum considered in the studies of stationary
models). Although it cannot be assumed that, in the real scenario,
the system adiabatically follows the vacuum of excitations from the
initial preparation, this simplification can serve to isolate one
of the basic mechanisms that can be responsible for the increase in
the density. Therefore, we calculate the mean value of $\hat{\rho}$
in the adiabatically evolved state $\left|0\right\rangle $. Using
Eqs. (\ref{eq:AdEXpQuan}) and (\ref{eq:NonAdexpan}), we find

\begin{eqnarray}
\left\langle 0\right|\rho_{0}^{ad}+\hat{\rho}_{1}^{ad}+\rho_{1}^{nad}\left|0\right\rangle  & = & \rho_{0}^{ad}+\rho_{1}^{nad}\sim\nonumber \\
 &  & \rho_{0}^{ad}(x,\boldsymbol{\Lambda}(t))\Bigl[1+\sum_{a}2\textrm{Re}\Bigl\{ B_{a}(t)\sigma_{a}(x,\boldsymbol{\Lambda}(t))+\nonumber \\
 &  & C_{a}(t)\nu_{a}(x,\boldsymbol{\Lambda}(t))\Bigr\}\Bigr],\label{eq:MeanVdens}
\end{eqnarray}
 where we have taken into account that, in parallel with the result
for a strictly static setup, the mean value of the quantum term is
zero in the vacuum state, i.e., 

\begin{equation}
\left\langle 0\right|\hat{\rho}_{1}^{ad}\left|0\right\rangle =0.\label{eq:MeanValDensAd}
\end{equation}
 The functions $\sigma_{a}(x,\boldsymbol{\Lambda}(t))$ and $\nu_{a}(x,\boldsymbol{\Lambda}(t))$
present in Eq. (\ref{eq:MeanVdens}) are the counterparts for $\boldsymbol{\mathcal{L}_{BdG}^{ad}}$
of the functions respectively given in Eq. (\ref{eq:staticComWaveFunc})
for the eigenmodes of $\boldsymbol{\mathcal{L}_{BdG}}$. 

It is important to emphasize that, as can be shown from the analysis
of Eqs. (\ref{eq:Bfunction}) and (\ref{eq:Cfunction}), the functional
forms of $B_{a}(t)$ and $C_{a}(t)$ depart from pure exponentials.
Namely, in Eq. (\ref{eq:Bfunction}), the time dependence is contained
not only in the factor $e^{-i\vartheta_{a}(t)}$, but also in the
integral of $e^{i\vartheta_{a}(t^{\prime})}$, in the adiabatic wavefunctions,
and, importantly, in the source term. Yet, assuming that the increasing
(decreasing) character of $B_{a}(t)$ $[C_{a}(t)]$ is robust against
nonadiabatic corrections, and, therefore, that, at sufficiently large
times, the terms that include $B_{a}(t)$ dominate, we can approximate
the total density by 

\begin{equation}
\left\langle 0\right|\hat{\rho}\left|0\right\rangle \sim\rho_{0}^{ad}(x,\boldsymbol{\Lambda}(t))\left[1+2\textrm{Re}\left\{ B_{a}(t)\sigma_{a}(x,\boldsymbol{\Lambda}(t))\right\} \right],\label{eq:MeanValDenLargeTim}
\end{equation}
where only the contribution of the most unstable growing term with
nonzero projection with the source has been kept. Hence, in agreement
with the experimental results, and, in contrast with the picture corresponding
to a stationary flow (see the Appendix), changes in the mean value
of the density are observed here. (It is worth recalling that the
variation in the density is a salient feature of the experimental
realization). Two components can be singled out in the spatial pattern
of those changes. The first one is the adiabatic wave function of
the background field, which enters Eq. (\ref{eq:MeanValDenLargeTim})
through $\rho_{0}^{ad}(x,\boldsymbol{\Lambda}(t))$. The second component
is the wave function of the most unstable mode that projects with
the nonadiabatic source term, which is incorporated in Eq. (\ref{eq:MeanValDenLargeTim})
via $\sigma_{a}(x,\boldsymbol{\Lambda}(t))$. It is the second component
that gives the peculiar character to the density evolution: because
of the dynamical instability, the effect of the nonadiabatic variation
of the system parameters is amplified; then, it can lead to drastic
changes in the density. Extrapolating these conclusions, derived in
a \emph{post-adiabatic} picture, to a general time-varying scenario,
we conjecture that the mere dynamical implementation of the black-hole
lasing configuration can provide the seed for an instability-determined
variation in the system density. 

Note that, at the considered order of approximation, the nonadiabatic
correction has a purely deterministic effect. Therefore, it is not
seen in the DCF if the mean value of the density is extracted in the
evaluation of the correlation, as it is done in the standard procedure.
Consequently, the DCF is merely determined by the solutions to the
homogeneous equation; namely, it reads 

\begin{eqnarray}
\left\langle 0\right|\hat{\rho}_{1}^{ad}(x,t)\hat{\rho}_{1}^{ad}(x^{\prime},t)\left|0\right\rangle  & \sim & \rho_{0}^{ad}(x,\boldsymbol{\Lambda}(t))\rho_{0}^{ad}(x^{\prime},\boldsymbol{\Lambda}(t))\times\nonumber \\
 &  & \sum_{a}e^{2\Delta_{a}(t)}\textrm{Re}\left\{ \sigma_{a}(x,\boldsymbol{\Lambda}(t))\sigma_{a}^{*}(x^{\prime},\boldsymbol{\Lambda}(t))\right\} ,\label{eq:DCFadiab}
\end{eqnarray}
where

\begin{equation}
\Delta_{a}(t)=\int_{0}^{t}\Gamma_{a}(\boldsymbol{\Lambda}(t^{\prime}))dt^{\prime}.\label{eq:delta}
\end{equation}
Therefore, in parallel with the results observed in the experiments,
we find that, in our toy model, the form of the time dependence of
the nonlocal correlation function, given by the exponential $e^{2\Delta_{a}(t)}$,
differs from that of the density, incorporated by the function $B_{a}(t)$
in Eq. (\ref{eq:MeanValDenLargeTim}). (We recall that an approximate
exponential behavior of experimental DCF is reported in Fig. 5 in
\cite{key-Steinhauer1}). In contrast, the presence of $\sigma_{a}(x,\boldsymbol{\Lambda}(t))$
in the obtained expressions for both observables, implies a similar
spatial form, also in agreement with the experimental findings. (The
analysis of the experimental results has actually uncovered a similar
pattern in the spatial dependence of the DCF and of the density \cite{key-Clark}).
In previous work on static models \cite{key-Coutant,key-Michel1,key-Carusotto4},
the wavefunctions of the unstable modes have been evaluated using
different approximate methods. Despite the departures of the practical
implementation from those simplified setups, a certain similarity
still exists between the model-proposed profiles of the velocity of
the flow and the speed of sound and those measured in the experiments.
Hence, the undulations observed in practice can be linked to the spatial
oscillations of the calculated theoretical wavefunctions.

\section{The effect of the instabilities on the classical noise}

Classical fluctuations are present in the practical arrangements for
BHL. The study of their implications is required: the identification
of differential effects of the instabilities on classical and quantum
fluctuations can be crucial to trace the presence of self-amplified
HR. A variety of forms of classical noise can be considered. Depending
on their characteristics, their effect on the analyzed observables
can vary. Let us exemplify this diversity of noisy responses by dealing
with two types of fluctuations potentially relevant to the experimental
realization \cite{key-Steinhauer1}.

\subsection{Noise in the system preparation}

The limited precision in the preparation of the initial state in the
experiments can be accounted for in our model by including classical
fluctuations $\delta\Phi^{cn}$ in the initial mean-field wavefunction.
For instance, shot-to-shot variations in the number of atoms can be
simulated in this form. This type of classical noise is incorporated
into our scheme via the two-component \emph{spinor} 

\begin{equation}
\boldsymbol{\delta\Psi^{cn}}\equiv\begin{pmatrix}\delta\Psi^{cn}\\
\delta\Psi^{cn*}
\end{pmatrix},\label{eq:NoiseSpinorPrep}
\end{equation}
 which is modified in each noise realization. Using the classical
version of Eq. (\ref{eq:AdEXpQuan}), $\boldsymbol{\delta\Psi^{cn}}$
is expanded in terms of the eigenmodes of $\boldsymbol{\mathcal{L}_{BdG}^{ad}}$.
Then, retaining only the contribution of the unstable modes, we write
for the evolved noisy perturbation
\begin{eqnarray}
\delta\Psi^{cn}(x,t) & \sim & e^{-\frac{i}{\hbar}\int_{0}^{t}\mu(\boldsymbol{\Lambda}(t^{\prime}))dt^{\prime}}\Phi_{0}^{ad}(x,\boldsymbol{\Lambda}(t))\times\nonumber \\
 &  & \sum_{a}\biggr[b_{a}e^{-i\vartheta_{a}(t)}\xi_{a}(x,\boldsymbol{\Lambda}(t))+c_{a}e^{-i\vartheta_{a}^{*}(t)}\psi_{a}(x,\boldsymbol{\Lambda}(t))+\nonumber \\
 &  & b_{a}^{*}e^{i\vartheta_{a}^{*}(t)}\eta_{a}^{*}(x,\boldsymbol{\Lambda}(t))+c_{a}^{*}e^{i\vartheta_{a}(t)}\zeta_{a}^{*}(x,\boldsymbol{\Lambda}(t))\biggl],\label{eq:NoisePrepExpan}
\end{eqnarray}
 where the coefficients $b_{a}$ and $c_{a}$ respectively denote
the projections of the noise \emph{spinor} on the modes $\boldsymbol{V_{a}}$
and $\boldsymbol{Z_{a}}$, i.e., they are given by 
\begin{eqnarray}
b_{a} & = & \left\langle \boldsymbol{V_{a}}\right|\left.\boldsymbol{\delta\Psi^{cn}}\right\rangle \nonumber \\
 & = & \int_{-\infty}^{\infty}dx\left[\Phi_{0}^{ad*}(x,\boldsymbol{\Lambda}(t))\xi_{a}^{*}(x,\boldsymbol{\Lambda}(t))\delta\Psi^{cn}(x)+\Phi_{0}^{ad}(x,\boldsymbol{\Lambda}(t))\eta_{a}^{*}(x,\boldsymbol{\Lambda}(t))\delta\Psi^{cn*}(x)\right]\label{eq:bCoeff-1}
\end{eqnarray}

\begin{eqnarray}
c_{a} & = & \left\langle \boldsymbol{Z_{a}}\right|\left.\boldsymbol{\delta\Psi^{cn}}\right\rangle \nonumber \\
 & = & \int_{-\infty}^{\infty}dx\left[\Phi_{0}^{ad*}(x,\boldsymbol{\Lambda}(t))\psi_{a}^{*}(x,\boldsymbol{\Lambda}(t))\delta\Psi^{cn}(x)+\Phi_{0}^{ad}(x,\boldsymbol{\Lambda}(t))\zeta_{a}^{*}(x,\boldsymbol{\Lambda}(t))\delta\Psi^{cn*}(x)\right].\label{eq:cCoeff-1}
\end{eqnarray}

Let us see how this kind of classical fluctuations can affect the
observables measured in the experiments. Eq. (\ref{eq:DensAd+NonAd})
for the density is replaced by $\hat{\rho}=\rho_{0}^{ad}+\hat{\rho}_{1}^{ad}+\rho_{1}^{nad}+\rho_{1}^{cn},$
where to first order in $\delta\Psi^{cn}(x,t)$, the classical-noise
contribution to the density is $\rho_{1}^{cn}=(\Psi_{0}^{ad}\delta\Psi^{cn*}+\textrm{c.c.})$.
Consequently, the variation in the mean value of $\hat{\rho}$ due
to the fluctuations is obtained as 

\begin{eqnarray}
\rho_{1}^{cn}(x,t) & \sim & \rho_{0}^{ad}(x,\boldsymbol{\Lambda}(t))\times\nonumber \\
 &  & \sum_{a}\textrm{Re}\left\{ b_{a}e^{-i\vartheta_{a}(t)}\sigma_{a}(x,\boldsymbol{\Lambda}(t))+c_{a}e^{-i\vartheta_{a}^{*}(t)}\nu_{a}(x,\boldsymbol{\Lambda}(t))\right\} ,\label{eq:Dens1NoisePrep}
\end{eqnarray}
 which, averaged over stochastic realizations $\left\langle \:\right\rangle _{nr}$,
gives

\begin{eqnarray}
\left\langle \rho_{1}^{cn}(x,t)\right\rangle _{nr} & \sim & \rho_{0}^{ad}(x,\boldsymbol{\Lambda}(t))\times\nonumber \\
 &  & \sum_{a}\textrm{Re}\left\{ \left\langle b_{a}\right\rangle _{nr}e^{-i\vartheta_{a}(t)}\sigma_{a}(x,\boldsymbol{\Lambda}(t))+\left\langle c_{a}\right\rangle _{nr}e^{-i\vartheta_{a}^{*}(t)}\nu_{a}(x,\boldsymbol{\Lambda}(t))\right\} .\label{eq:AverDens1NoisP}
\end{eqnarray}
 Then, if the noise has zero mean value, i.e., if $\left\langle \delta\Psi^{cn}\right\rangle _{nr}=0$,
which, in turn, implies $\left\langle b_{a}\right\rangle _{nr}=\left\langle c_{a}\right\rangle _{nr}=0$,
it follows that 
\begin{equation}
\left\langle \rho_{1}^{cn}(x,t)\right\rangle _{nr}=0,\label{eq:zeroValueDensNoiP}
\end{equation}
 which parallels the result obtained for quantum fluctuations (see
Eq. (\ref{eq:MeanValDensAd})). Additionally, the contribution of
noise to the DCF is straightforwardly obtained if the existence of
a clearly dominating unstable mode, i.e., of a mode with a growing
rate $\Gamma_{a}$ much larger than the others, is assumed. In that
case, we find

\begin{eqnarray}
\left\langle \rho_{1}^{cn}(x,t)\rho_{1}^{cn}(x^{\prime},t)\right\rangle _{nr} & \sim & \rho_{0}^{ad}(x,\boldsymbol{\Lambda}(t))\rho_{0}^{ad}(x^{\prime},\boldsymbol{\Lambda}(t))\times\nonumber \\
 &  & \left\langle \left|b_{a}\right|^{2}\right\rangle _{nr}e^{2\Delta_{a}(t)}\textrm{Re}\left\{ \sigma_{a}(x,\boldsymbol{\Lambda}(t))\sigma_{a}^{*}(x^{\prime},\boldsymbol{\Lambda}(t))\right\} ,\label{eq:DCFnoiseP}
\end{eqnarray}
 where have retained only the secular (nonoscillating) terms. Hence,
the pattern is given by the form of the wavefunction of the most unstable
mode. Again, this is an analog of the result corresponding to quantum
noise, as can be shown by retaining only the dominant term in Eq.
(\ref{eq:DCFadiab}). Therefore, in the considered temporal range,
namely, for times sufficiently large for having the dynamics determined
by a clearly dominant unstable mode, there are not qualitative differences
between the effects of quantum fluctuations and (classical) noise
in the system preparation.

\subsection{Technical noise}

We consider now fluctuations in the elements that constitute the practical
arrangement. For instance, let us deal with stochastic variations
in the realization of the optical step-potential used to implement
the two-horizon configuration. In this case, the different noise realizations
can be regarded as implementations of diverse setups. Therefore, a
different set of eigenmodes is applicable to each experimental run.
Additionally, there is effective noise in the preparation: the difference
between the prepared wave function and the stationary solution of
the GP equation for the actually realized set of parameters is taken
as a stochastic perturbation. The expansions given by Eqs. (\ref{eq:NoisePrepExpan}),
(\ref{eq:bCoeff-1}), and (\ref{eq:cCoeff-1}) are still applicable.
Furthermore, the density is  given by Eq. (\ref{eq:Dens1NoisePrep})
if only the contribution of the unstable modes is retained. The differences
with the effect of shot-to-shot noise become evident when the averages
over noise realizations are carried out. Namely, the average of density
outputs, keeping only the increasing terms, is now given by 

\begin{equation}
\left\langle \rho_{1}^{cn}(x,t)\right\rangle _{nr}\sim\rho_{0}^{ad}(x,\boldsymbol{\Lambda}(t))\sum_{a}\textrm{Re}\left\{ \left\langle b_{a}e^{-i\vartheta_{a}(t)}\sigma_{a}(x,\boldsymbol{\Lambda}(t))\right\rangle _{nr}\right\} ,\label{eq:TechAverDens}
\end{equation}
 where $b_{a}$ is the projection of the effective noise in the preparation
on the mode $\boldsymbol{V_{a}}$. It is important to realize that,
even when the fluctuations have zero average over realizations, they
can change the density mean value. Actually, as each realization implies
not only a different value of $b_{a}$, but also of $\sigma_{a}(x,\boldsymbol{\Lambda}(t))$,
the average of density outputs does not have to be zero, i.e., in
general, 
\begin{equation}
\left\langle \rho_{1}^{cn}(x,t)\right\rangle _{nr}\neq0.\label{eq:NonZeroTechAver}
\end{equation}
 Moreover, again, for times sufficiently large for having the evolution
determined by only a dominant growing rate $\Gamma_{a}$, we have
for the DCF

\begin{eqnarray}
\left\langle \rho_{1}^{cn}(x,t)\rho_{1}^{cn}(x^{\prime},t)\right\rangle _{nr} & \sim & \rho_{0}^{ad}(x,\boldsymbol{\Lambda}(t))\rho_{0}^{ad}(x^{\prime},\boldsymbol{\Lambda}(t))\times\nonumber \\
 &  & \left\langle \left|b_{a}\right|^{2}e^{2\Delta_{a}(t)}\textrm{Re}\left\{ \sigma_{a}(x,\boldsymbol{\Lambda}(t))\sigma_{a}^{*}(x^{\prime},\boldsymbol{\Lambda}(t))\right\} \right\rangle _{nr},\label{eq:techDCF}
\end{eqnarray}
 where $\Delta_{a}$, and $\sigma_{a}(x,\boldsymbol{\Lambda}(t))$
are characteristics of the most unstable mode for each realization
of system parameters. Note that, in this case, as the pattern incorporates
an average over wave-functions corresponding to the differently realized
set of parameters, the form of the density correlation function differs
from that obtained in the previously analyzed cases, i.e., for quantum
fluctuations and for noise in the system preparation. In particular,
in the present case, since slightly different wave-functions are being
averaged, the contrast of the pattern lines can be expected to weaken.

\subsection{Interference effects in the system response to classical noise}

In the above analysis of the DCF, we have considered the case where
there is one unstable mode whose frequency has an imaginary part $\Gamma_{a}$
much larger than those of the rest of modes. Then, at times sufficiently
large, the DCF is determined by the characteristics of that dominant
mode. No spectral structure is then observed. Let us evaluate now
the appearance of qualitative differences in the situation corresponding
to having, at least, two unstable modes whose frequencies have imaginary
parts, $\Gamma_{a1}$ and $\Gamma_{a2}$, of the same magnitude. The
analysis is not specific to the adiabatic scenario. In fact, it also
applies to a stationary setup. To emphasize its generality, we suppress
the references to adiabaticity in the notation. 

For quantum noise, when there are two predominant unstable modes,
we obtain from Eq. (\ref{eq:staticDCFdomin})

\begin{equation}
\left\langle 0\right|\hat{\rho_{1}}(x,t)\hat{\rho_{1}}(x^{\prime},t)\left|0\right\rangle \sim e^{2\Gamma_{a1}t}\textrm{Re}\left\{ \sigma_{a1}(x)\sigma_{a1}^{*}(x^{\prime})\right\} +e^{2\Gamma_{a2}t}\textrm{Re}\left\{ \sigma_{a2}(x)\sigma_{a2}^{*}(x^{\prime})\right\} .\label{eq:2modesQuan}
\end{equation}
Hence, no interferences effects emerge in the mean value of the DCF
if the system can be assumed to be prepared in the vacuum state. (Here,
it is worth pointing out that the experimental conditions correspond
to a more complex situation. Indeed, since, in the practical realization,
the system departs from a stable regime and enters the instability
region through the variation in the external potential, the adiabatic
following of the vacuum state, assumed in our toy model, does not
hold. Therefore, one cannot discard the emergence of spectral structure
rooted in the nontrivial evolution of the initial state).

In contrast with the prediction for quantum fluctuations, for (classical)
noise in the system preparation, we find 

\begin{eqnarray}
\left\langle \rho_{1}(x,t)\rho_{1}(x^{\prime},t)\right\rangle _{nr} & \sim & \left\langle \left|b_{a1}\right|^{2}\right\rangle _{nr}e^{2\Gamma_{a1}t}\textrm{Re}\left\{ \sigma_{a1}(x)\sigma_{a1}^{*}(x^{\prime})\right\} +\nonumber \\
 &  & \left\langle \left|b_{a2}\right|^{2}\right\rangle _{nr}e^{2\Gamma_{a2}t}\textrm{Re}\left\{ \sigma_{a2}(x)\sigma_{a2}^{*}(x^{\prime})\right\} +\nonumber \\
 &  & e^{(\Gamma_{a1}+\Gamma_{a2})t}\textrm{Re}\left\{ e^{-i(\omega_{a1}-\omega_{a2})t}\left\langle b_{a1}b_{a2}^{*}\right\rangle _{nr}[\sigma_{a1}(x)\sigma_{a2}^{*}(x^{\prime})+\sigma_{a1}(x^{\prime})\sigma_{a2}^{*}(x)]\right\} ,\label{eq:2modesNoiseP}
\end{eqnarray}
 where an oscillating term with a frequency given by the difference
between the real parts of the two considered eigenvalues is apparent.
The counterpart expression for technical noise reads

\begin{eqnarray}
\left\langle \rho_{1}(x,t)\rho_{1}(x^{\prime},t)\right\rangle _{nr} & \sim & \biggl\langle\left|b_{a1}\right|^{2}e^{2\Gamma_{a1}t}\textrm{Re}\left\{ \sigma_{a1}(x)\sigma_{a1}^{*}(x^{\prime})\right\} \biggr\rangle_{nr}+\nonumber \\
 &  & \biggl\langle\left|b_{a2}\right|^{2}e^{2\Gamma_{a2}t}\textrm{Re}\left\{ \sigma_{a2}(x)\sigma_{a2}^{*}(x^{\prime})\right\} \biggr\rangle_{nr}+\nonumber \\
 &  & \biggl\langle e^{(\Gamma_{a1}+\Gamma_{a2})t}\textrm{Re}\left\{ e^{-i(\omega_{a1}-\omega_{a2})t}b_{a1}b_{a2}^{*}[\sigma_{a1}(x)\sigma_{a2}^{*}(x^{\prime})+\sigma_{a1}(x^{\prime})\sigma_{a2}^{*}(x)]\right\} \biggr\rangle_{nr},\label{eq:2modesTech}
\end{eqnarray}
 where, again, oscillations are observed. It is then concluded that,
for the two types of classical noise previously considered, interferences
appear in the DCF. In the evaluation of the practical importance of
these results, one must keep in mind that the approximation of retaining
only the most unstable mode in the description becomes worse as the
time of observation is reduced: at shorter times, more unstable modes
have a non negligible role in the dynamics. Given the time limitations
to maintain the two-horizon scheme in practice, the interference effects
cannot be neglected. Then, in the analysis of the spectral characteristics
of the DCF observed in the experiments, the relevance of more than
one unstable mode to the dynamics, and, as a consequence, the resulting
interference effects on classical fluctuations, must be taken into
account. As previously indicated, one can also contemplate the potential
emergence of spectral structure in the quantum scenario when the adiabatic
following of the vacuum state does not hold: the breakdown of the
adiabatic approximation implies the population of higher modes, and,
consequently, the appearance of nontrivial spectral features. An open
question is how the quantum fluctuations in the preparation are affected
by the system entering the instability region. In this context of
discriminating the classical and quantum origin of the findings, it
is worth recalling the applicability of the measurement of entanglement
 as an unambiguous signature for quantum behavior \cite{key-Steinhauer2}.

\section{Numerical results}

In order to confirm the predictions of the above analytical study,
we present in this section a numerical simulation of the referred
experiments \cite{key-Steinhauer1}. We depart from the three-dimensional
GP equation where we incorporate the time variation of the trapping
potential implemented in the practical setup. The dimensional reduction
allowed by the strong confinement in the directions transversal to
the step displacement leads to the Non-Polynomial Schrödinger Equation
(NPSE) for the longitudinal coordinate. The system is considered to
be prepared in the ground state corresponding to the initial confining
potential. To emulate the time dependent potential we use the functional
form and parameters, in particular the step velocity $v_{s}$, given
in Ref. \cite{key-Carusotto1}. The NPSE is solved using split-operator
and fast Fourier-transform techniques. Results for the sound speed
and for the fluid velocity at two different times are presented in
Fig. 1.

\begin{figure}[H]
\centerline{\includegraphics{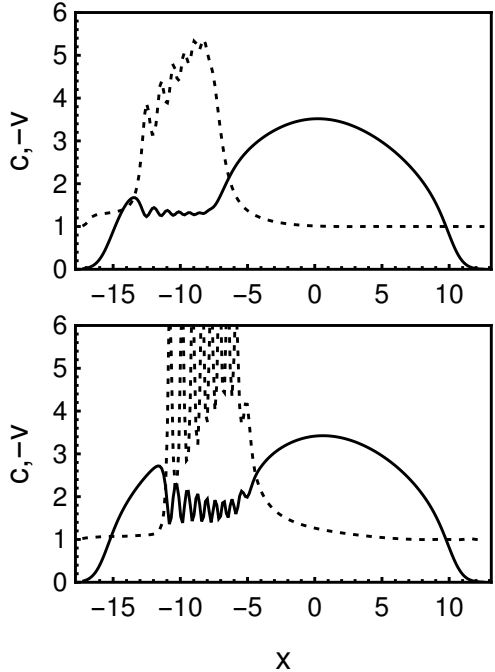}}\caption{The fluid velocity in the step frame $-v(x)$ (dashed line) and the
sound speed $c(x)$ (solid line) as given by the numerical simulation
of the experiment of Ref. \cite{key-Steinhauer1} at two different
times $t_{1}$ (upper panel), and $t_{2}$ (lower panel) with $t_{2}-t_{1}=0.1$
and arbitrary time origin. The system parameters are the same as those
used in Ref. \cite{key-Carusotto1}. (We have used reduced dimensionless
units which correspond to work with $\hbar=1$, $m=1$, and $v_{s}=1$).}
\end{figure}

The similarity with the experimental findings is evident. In particular,
a conspicuous feature of the practical realization of Ref. \cite{key-Steinhauer1}
is reproduced: although an adiabatic approximation is not applicable,
a relatively slow variation of the observables is apparent once the
white horizon is formed. This aspect of the dynamics has been a key
element in the design of our toy model. Indeed, one can think of using
a postadiabatic scenario, with an appropriately chosen set of slowly
varying parameters, to obtain insight into the observed behavior.
Now, focusing on that time regime, we will show that, as analytically
predicted, any departure from adiabaticity can become a seed for instability.
It is worth recalling that the general objective of our approach has
been to establish a link between previous theoretical studies of instability,
carried out in static models, and the nonstationary experimental realization.
Here, in order to apply the theory to the practical setup, we must
turn to a static parallel system where the characterization of the
modes can be feasible. Moreover, nonstationarity must be introduced
in that analogue system in a form that emulates the detected behavior.
Accordingly, we proceed as follows:

i) We employ a monodimensional stationary model used in former work
on static BHL settings \cite{key-Finazzi1}. In it, the sound speed
$c(x)$ and the fluid velocity $v(x)$ are given by the equations 

\begin{equation}
c(x)+v(x)=D_{c}\tanh\left(\frac{\kappa_{W}(x+L)}{D_{c}}\right)\tanh\left(\frac{\kappa_{B}(x-L)}{D_{c}}\right)\label{eq:c+v}
\end{equation}

\begin{equation}
c(x)=c_{H}+(1-q)[c(x)+v(x)]\label{eq:c(x)}
\end{equation}

\begin{equation}
v(x)=-c_{H}+q[c(x)+v(x)].\label{eq:v(x)}
\end{equation}
 In this framework, the white horizon is located at $-L$ with \emph{surface
gravity} $-c_{H}\kappa_{W}$, and the black horizon is placed at $L$
with \emph{surface gravity} $c_{H}\kappa_{B}$. {[}$c_{H}$ denotes
the sound speed at the horizons, $D_{c}/c_{H}$ gives the extension
of the range (close to the horizons) where both velocities are not
flat, and $q$ specifies which part of the sum $c(x)+v(x)$ corresponds
to each velocity{]}. The connection of this model with the practical
setup is evident in Fig. 2 (upper panel), where we depict $c(x)$
and $v(x)$ as given by Eqs. (45), (46), and (47). Actually, the characterization
of the modes in this system provided interesting clues (\cite{key-Finazzi1})
to modify an early implementation of the BHL setup (\cite{key-Steinhauer3})
in order to facilitate the detection of the amplification. Namely,
a change of parameters was proposed to enhance the instability, i.e.,
to increase the imaginary part of the relevant eigenfrequencies.

ii) By time varying some of the parameters of the model, we incorporate
nonstationarity in the setup. Since, in the experimental curves, both,
$c$ and $v$, are seen to increase with time, it is appropriate to
introduce nonstationarity through the time variation of $c_{H}$.
(The implications of changes in other parameters will be discussed
further on). Specifically, we consider that $c_{H}$ is modified according
to 
\begin{equation}
c_{H}(t)=c_{H_{0}}+\alpha t.\label{eq:cvtimevar}
\end{equation}
 Our procedure to describe the dynamics resulting from the variation
of $c_{H}$ consists in obtaining first the confining potential $V(x)$
and the effective interaction strength $g(x)$ which lead to a ground-state
solution of the GP equation with the characteristics given by Eqs.
(\ref{eq:c+v}), (\ref{eq:c(x)}), (\ref{eq:v(x)}) with $c_{H}=c_{H_{0}}$
\cite{key-Finazzi1}. Additionally, the (thus derived) potential and
interaction strength are made to incorporate the time-varying $c_{H}(t)$.
Then, taking as initial preparation the wave function determined by
Eqs. (\ref{eq:c+v}), (\ref{eq:c(x)}), (\ref{eq:v(x)}) with $c_{H}=c_{H_{0}}$,
we solve the GP equation with the time-dependent potential and interaction
strength, $V(x,t)$ and $g(x,t)$. We have chosen system parameters
appropriate to emulate the practical setup. They correspond to the
existence of unstable modes. We intend to confirm that, as shown in
Sec. III, {[}see Eq. (\ref{eq:bCoeff-1}){]}, the perturbation of
the system, given by the projection of the source term on the modes,
is amplified provided that any of the involved modes is unstable,
and, that, in turn, it leads to a significant variation in the density.
The magnitude of nonadiabaticity is controlled with the parameter
$\alpha$. Results for the sound speed and fluid velocity at two different
times are presented in Fig. 2 (middle and lower panels). (The thiner
lines stand for the strictly adiabatic regime. The thicker lines correspond
to $\alpha=0.02$). The differences between the results for the two
regimes of time variation show that, as predicted by our analytical
study, a departure from adiabaticity can become a seed for activating
the instability of the system. It is worth emphasizing the time growing
character of those differences. 

\begin{figure}[H]
\centerline{\includegraphics{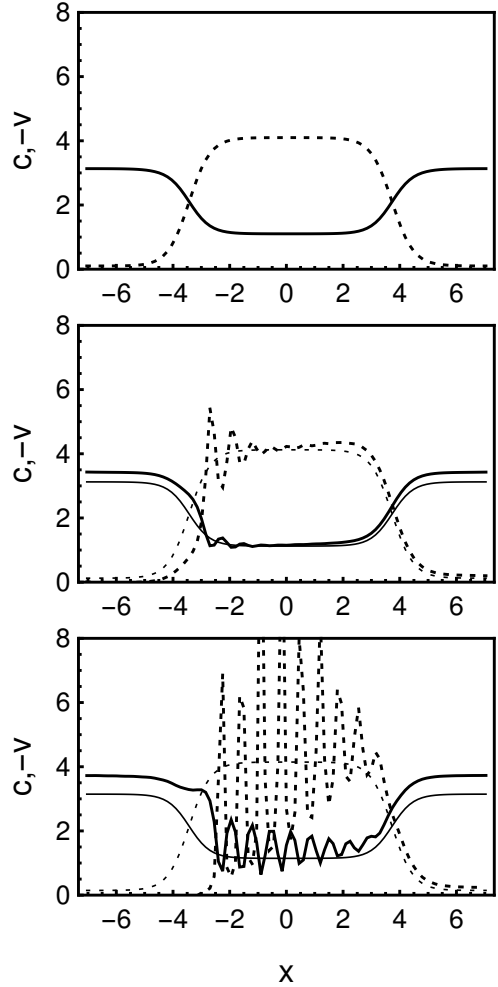}}\caption{The fluid velocity $v(x)$ (dashed line) and the sound speed $c(x)$
(solid line) as given by the model defined by Eqs. (45), (46), (47),
and (47) at three different times $t=0$ (upper panel), $t=0.06$
(middle panel), and $t=0.12$ (lower panel). Note that the upper panel
corresponds to the (permanent) velocity profiles in the stationary
system. In the middle and lower panels, the thiner lines represent
the strictly adiabatic regime and the thicker lines correspond to
$\alpha=0.02$. The rest of model parameters are: $c_{H_{0}}=2.1$,
$L=3.58$, $D_{c}=3$, $\kappa_{W}=\kappa_{B}=4$, and $q=2/3$. (We
have used reduced dimensionless units which correspond to work with
$\hbar=1$, $m=1$, and $v_{s}=1$).}
\end{figure}

The properties of the modes in the static system, i.e., in the model
with frozen parameters, give further insight into the observed behavior.
The use in a former study of a semiclassical approximation to analyze
them has provided useful information on the real and imaginary parts
of the eigenfrequencies and on the associated eigenfunctions. That
study focused on the dependence of the mode properties on $L$. The
application of that approach, adapted to the present scenario of fixed
$L$ and slightly varying $c_{H}$, allows us to test our predictions
on the role of the modes in the emergent dynamics. We have produced
two different proofs of consistency. First, from the number of oscillations
in the density, we have estimated the wavelength of the leading unstable
mode. Then, from the approximately known dependence of the modes on
$L$, we have evaluated the separation between horizons that corresponds
to having that mode as dominant. (From the application of the quantization
rule of Bohr-Sommerfeld, we can assume that no changes take place
in the number of unstable modes as $c_{H}$ is varied in the considered
time interval). We have found that the required length has the same
magnitude as that corresponding to the experiment. Second, from our
characterization of the modes, we have been able to alter the system
parameters to reduce in a controlled way the number of oscillations
in the density. The results are shown in Fig. 3. (It must be noticed
that to obtain the curves in Fig. 3 we have kept the former model
parameters and have altered the reduced units to work with $\hbar=2$). 

\begin{figure}[H]
\centerline{\includegraphics{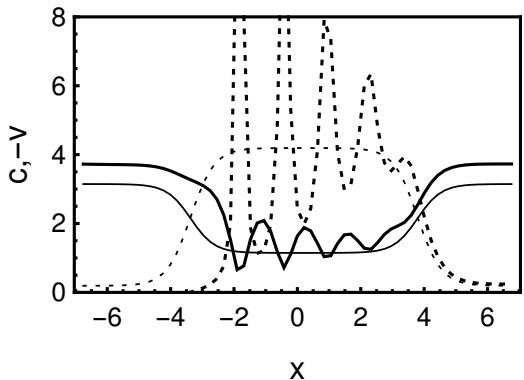}}\caption{The fluid velocity $v(x)$ and the sound speed $c(x)$ as given by
the model defined by Eqs. (45), (46), (47), and (47) at $t=0.12$.
The thiner lines represent the strictly adiabatic regime and the thicker
lines correspond to $\alpha=0.02$. The model parameters are the same
as those given in Fig 2. (We have used reduced dimensionless units,
different from those in Fig. 2, which correspond to work with $\hbar=2$,
$m=1$, and $v_{s}=1$).}
\end{figure}

It is apparent that despite the restrictions on its applicability,
the study offers a framework where a certain degree of control over
the system can be achieved. Advances in the design of optimized setups
can also be expected from the study of the effect of the variation
of other parameters. That analysis requires the characterization of
the dependence of the modes on those parameters. Indeed, in order
to predict the emergence of instability, the overlapping of the source
term with the mode eigenfunctions must be evaluated. The same argument
is applicable to the evaluation of the effect of other, deterministic
or noisy, perturbation. 

In summary, the results of our numerical study confirm the validity
of the picture given by our analytical approach. We emphasize that
the mere use of terms like instability or unstable modes, which is
frequent in the analysis of the experimental findings and which has
been applied in the design of the experimental setup, implies assuming
that an adiabatic scenario can give a useful ground for describing
the essential physics of the system. Our model has given support to
that use.

\section{Concluding remarks}

The non-stationary character of the implementation of BHL has been
shown to imply departures from the behavior predicted for a static
configuration. Specifically, the study of a post-adiabatic model has
uncovered differential aspects in the role of the dynamical instabilities.
Apart from leading to the amplification of quantum fluctuations, as
in a static setup, the instabilities induce nontrivial variations
in the density. The mechanism responsible for the growth of the density
mean-value incorporates different elements. First, the variation in
the adiabatic wave function of the substrate leads to changes in the
density, as expected also in a stable configuration. Second, specific
to the presence of instabilities is that the non-adiabatic corrections
get amplified provided that they have nonzero projection on the unstable
modes. Hence, the density grows because the nonstationary deterministic
seed is amplified via the unstable modes. It is this combination of
factors that makes the form of the time dependence of the density
to differ from that of the nonlocal DCF, which simply corresponds
to the exponential growth associated with the instabilities. On the
other hand, the similarity between the spatial patterns of both, the
density and the DCH, is basically due to the dependence of both observables
on the wave-functions of the dominant unstable modes. The generality
of the identified mechanism provides a certain predictive power on
the implications of the different system components. For instance,
although, in our model, the set of variable parameters $\boldsymbol{\Lambda}(t)$
has been assumed to enter the system through the external potential,
the consequences of changes in the interaction strength could equally
be evaluated. 

We have not found qualitative differences between the effects of quantum
noise and (classical) fluctuations in the preparation when only one
unstable mode dominates the dynamics. This analogy does not persist
in the case of technical noise: changes in the density and in the
DCF can take place. Furthermore, differential effects of the instabilities
on quantum and classical noise are apparent when more than one unstable
mode intervene. Namely, for quantum fluctuations from the (adiabatically
evolved) vacuum state, no interference effects appear in the DCF.
On the contrary, for the two types of classical noise previously considered,
oscillations, and, therefore, spectral structure, are apparent in
the DCF. 

A comment on the limitations of our model is pertinent. Given the
presence of instabilities in the system, the use of the adiabatic
approximation has required assuming restrictive conditions \cite{key-Anglin}.
In this sense, we have opted for working in a temporal range where
the applicability of our approach is guaranteed, and, still, where
valuable information on the physical mechanisms specific to non-stationarity
could be extracted. Whereas we have dealt with non-adiabatic corrections
to the classical field $\Phi_{0}^{ad}(x,\boldsymbol{\Lambda}(t))$,
the adiabatic following of, both, the vacuum state and the eigenstates
of the operator $\boldsymbol{\mathcal{L}_{BdG}^{ad}}$, has been assumed.
As our approach put the focus on a temporal range subsequent to the
formation of the two-horizon scheme, it does do not allow us to assess
the potential transition of the HR from spontaneous to self-amplifying.
A more complete description, which incorporates the breakdown of the
adiabatic approximation, and, in particular, the nontrivial evolution
of the quantum fluctuations from the initial state, can be expected
to account for more complex spectral characteristics. In spite of
being a drastic simplification of the experimental setup, our picture
is sufficiently accurate to provide arguments useful in the discussion
of the findings. 

\appendix*

\section{DESCRIPTION OF THE STATIC SCENARIO}

\subsection{Characterization of the eigenmodes of the Bogoliubov-De Gennes operator
$\boldsymbol{\mathcal{L}_{BdG}}$}

To describe the elementary excitations of the condensate in a static
scenario, i.e., of the considered system for fixed values of the parameters
$\boldsymbol{\Lambda}(t)$, we simply rewrite the ansatz in Eq. (\ref{eq:BasicAnsatz}1)
as 
\begin{equation}
\hat{\Psi}=(\Phi_{0}+\hat{\delta\Phi})e^{-i\mu t/\hbar},\label{A1:staticAnsatz}
\end{equation}
 where $\mu$ denotes the chemical potential. Then, it is found that 

\begin{equation}
H_{GP}\Phi_{0}(x)=\mu\Phi_{0}(x),\label{eq:staticGPequ}
\end{equation}
with $H_{GP}$ representing the GP Hamiltonian, i.e., 

\begin{equation}
H_{GP}=-\frac{\hbar^{2}}{2m}\frac{d^{2}}{dx^{2}}+V_{ex}(x)+g\left|\Phi_{0}(x)\right|^{2}.\label{eq:staticBGHamil}
\end{equation}
{[}The (fixed) confining potential is denoted as $V_{ex}(x)${]}.
Assuming that $\Phi_{0}$ and $\mu$ are known, the evolution  of
the quantum contribution $\hat{\delta\Psi}$ can be formally analyzed
\cite{key-Finazzi1,key-Leonhardt}. In a linear approximation, the
dynamics of the excitations are given by the BdG approach. Following
the standard procedure, we define the two-component field $\hat{\boldsymbol{W}}\equiv\begin{pmatrix}\hat{\delta\Phi}\\
\hat{\delta\Phi}^{\dagger}
\end{pmatrix}$, and work initially with its classical version $\boldsymbol{W}\equiv\begin{pmatrix}\delta\Phi\\
\delta\Phi^{*}
\end{pmatrix}$. Using Eqs. (\ref{A1:staticAnsatz}), (\ref{eq:staticGPequ}), and
(\ref{eq:staticBGHamil}), we find 

\begin{equation}
\frac{\partial\boldsymbol{W}}{\partial t}=-\frac{i}{\hbar}\boldsymbol{\mathcal{L}_{BdG}}\boldsymbol{W},\label{eq:statBdGequ}
\end{equation}
where the BdG operator $\boldsymbol{\mathcal{L}_{BdG}}$ is given
by 

\begin{equation}
\boldsymbol{\mathcal{L}_{BdG}}\equiv\begin{pmatrix}H_{GP}-\mu+2g\left|\Phi_{0}\right|^{2} & g\Phi_{0}^{2}(x)\\
-g\Phi_{0}^{*2}(x) & -\left(H_{GP}-\mu+2g\left|\Phi_{0}\right|^{2}\right)
\end{pmatrix}.\label{eq:staticBdGoper}
\end{equation}
 The spectrum of $\boldsymbol{\mathcal{L}_{BdG}}$ has been analyzed
in previous research. From the non-Hermitian character of $\boldsymbol{\mathcal{L}_{BdG}}$,
it follows that its eigenvalues can be real or complex. The complex
eigenfrequencies constitute a discrete spectrum, and, given the characteristics
of $\boldsymbol{\mathcal{L}_{BdG}}$, appear in pairs, which will
be denoted as $\lambda_{a}$ and $\lambda_{a}^{*}$ ($\lambda_{a}=\omega_{a}+i\Gamma_{a}$).
Without loss of generality, we assume $\Gamma_{a}>0$. The respective
associated eigenmodes can be written as 

\begin{equation}
\boldsymbol{V_{a}}(x)=\begin{pmatrix}\Phi_{0}(x)\xi_{a}(x)\\
\Phi_{0}^{*}(x)\eta_{a}(x)
\end{pmatrix},\qquad\boldsymbol{Z_{a}}(x,)=\begin{pmatrix}\Phi_{0}(x)\psi_{a}(x)\\
\Phi_{0}^{*}(x)\zeta_{a}(x)
\end{pmatrix},\label{eq:staticeigenm}
\end{equation}
The imaginary parts of the eigenvalues lead to instabilities which
are responsible for the lasing effect. The growth rate of the perturbations
is given by $\Gamma_{a}$. Additionally, the real eigenfrequencies
are shown to form a continuous spectrum. They will be denoted as $\omega$,
and the associated eigenmodes can be expressed as  

\begin{equation}
\boldsymbol{W_{\omega}^{\alpha}}(x)=\begin{pmatrix}\Phi_{0}(x)\phi_{\omega}^{\alpha}(x)\\
\Phi_{0}^{*}(x)\varphi_{\omega}^{\alpha}(x)
\end{pmatrix},\label{eq:staticConteigen}
\end{equation}
 The variable $\alpha$ accounts for possible degeneracies. The field
$\hat{\boldsymbol{W}}$ can be expanded as $\hat{\boldsymbol{W}}=\sum_{n}\boldsymbol{W_{n}}\hat{a}_{n}+\bar{\boldsymbol{W}}_{\boldsymbol{n}}\hat{a}_{n}^{\dagger}$,
where $\boldsymbol{W_{n}}$ denotes the different eigenmodes (both
of the continuum and discrete spectrum) and $\bar{\boldsymbol{W}}_{\boldsymbol{n}}$
stands for the conjugate spinor of $\boldsymbol{W_{n}}$, which, using
the Pauli matrix $\sigma_{x}$, is defined as $\bar{\boldsymbol{W}}_{\boldsymbol{n}}=\sigma_{x}\boldsymbol{W_{n}^{*}}$.
Moreover, $\hat{a}_{n}$ is a compact notation for the two kinds of
operators potentially present in the description: it incorporates
the operators $\hat{a}_{\omega}^{\alpha}$ corresponding to the continuous
set of modes, and it also stands for the operators $\hat{b}_{a}$
and $\hat{c}_{a}$ linked to the discrete set. The commutators are
 $\left[\hat{a}_{\omega}^{\alpha},\hat{a}_{\omega^{\prime}}^{\alpha^{\prime}\dagger}\right]=\delta_{\alpha,\alpha^{\prime}}\delta(\omega-\omega^{\prime})$,
and $\left[\hat{b}_{a},\hat{c}_{a^{\prime}}^{\dagger}\right]=i\delta_{a,a^{\prime}}$,
vanishing all the others. As opposed to $\hat{a}_{\omega}^{\alpha}$,
$\hat{b}_{a}$ and $\hat{c}_{a}$ are not annihilation operators \cite{key-Finazzi1}.

It is shown that, in a general case {[}i.e., when $\boldsymbol{\mathcal{L}_{BdG}}$
has a continuous set of modes $\boldsymbol{W_{\omega}^{\alpha}}$
and a discrete set of (unstable) modes $\boldsymbol{V_{a}}$ and $\boldsymbol{Z_{a}}${]},
the expansion of the field operator $\hat{\delta\Psi}(x,t)=e^{-i\mu t/\hbar}\hat{\delta\Phi}(x,t)$
is given by 

\begin{eqnarray}
\hat{\delta\Psi}(x,t) & = & e^{-i\mu t/\hbar}\Phi_{0}(x)\biggl\{\int d\omega\sum_{\alpha}\left(e^{-i\omega t}\phi_{\omega}^{\alpha}(x)\hat{a}_{\omega}^{\alpha}+e^{-i\omega t}\varphi_{\omega}^{\alpha}(x)\hat{a}_{\omega}^{\alpha\dagger}\right)+\nonumber \\
 &  & \sum_{a}\left(e^{-i\lambda_{a}t}\xi_{a}(x)\hat{b}_{a}+e^{-i\lambda_{a}^{*}t}\psi_{a}(x)\hat{c}_{a}+e^{i\lambda_{a}^{*}t}\eta_{a}^{*}(x)\hat{b}_{a}^{\dagger}+e^{i\lambda_{a}t}\zeta_{a}^{*}(x)\hat{c}_{a}^{\dagger}\right)\biggr\}.\label{eq:staticExpan}
\end{eqnarray}

\subsection{The density and the density-density correlation function }

To first order in $\hat{\delta\Psi}(x,t)$, we have for the density
$\hat{\rho}=\rho_{0}+(\Phi_{0}\hat{\delta\Phi}^{\dagger}+\textrm{h.c.})\equiv\rho_{0}+\hat{\rho}_{1}$,
($\rho_{0}(x)=\left|\Phi_{0}(x)\right|^{2}$). As we intend to account
for Hawking radiation, we consider the system prepared in the vacuum
state $\left|0\right\rangle $. It is worth clarifying that $\left|0\right\rangle $
refers to the vacuum of the real annihilation operators $\hat{d}_{a+}$
and $\hat{d}_{a-}$ defined in the standard form $\hat{d}_{a+}=\frac{\hat{b}_{a}+i\hat{c}_{a}}{\sqrt{2}}$,
and $\hat{d}_{a-}=\frac{\hat{b}_{a}^{\dagger}+i\hat{c}_{a}^{\dagger}}{\sqrt{2}}.$
Introducing those changes into the expansion in Eq. (\ref{eq:staticExpan}),
we find $\left\langle 0\right|\hat{\rho}_{1}\left|0\right\rangle =0.$
Additionally, for the nonlocal DCF, we have

\begin{eqnarray}
\left\langle 0\right|\hat{\rho}_{1}(x,t)\hat{\rho}_{1}(x^{\prime},t)\left|0\right\rangle  & \sim & \rho_{0}(x)\rho_{0}(x^{\prime})\times\nonumber \\
 &  & \sum_{a}\left(e^{2\Gamma_{a}t}\textrm{Re}\left\{ \sigma_{a}(x)\sigma_{a}^{*}(x^{\prime})\right\} +e^{-2\Gamma_{a}t}\textrm{Re}\left\{ \nu_{a}(x)\nu_{a}^{*}(x^{\prime})\right\} \right),\label{eq:staticDCF}
\end{eqnarray}
 where we have retained only the contribution of the unstable modes.
The functions 
\begin{equation}
\sigma_{a}(x)=\xi_{a}(x)+\eta_{a}(x),\:\textrm{and}\qquad\nu_{a}(x)=\psi_{a}(x)+\zeta_{a}(x),\label{eq:staticComWaveFunc}
\end{equation}
 respectively correspond to the increasing and decreasing modes. Keeping
only the increasing terms (recall that $\Gamma_{a}>0$), we obtain
\cite{key-Finazzi1} 

\begin{equation}
\left\langle 0\right|\hat{\rho}_{1}(x,t)\hat{\rho}_{1}(x^{\prime},t)\left|0\right\rangle \sim\rho_{0}(x)\rho_{0}(x^{\prime})\sum_{a}e^{2\Gamma_{a}t}\textrm{Re}\left\{ \sigma_{a}(x)\sigma_{a}^{*}(x^{\prime})\right\} .\label{eq:staticDCFdomin}
\end{equation}
Furthermore, if there is one unstable mode whose frequency has an
imaginary part $\Gamma_{a}$ much larger than that of any other mode,
we can, for sufficiently large times, keep only the contribution of
that mode in Eq. (\ref{eq:staticDCFdomin}). An exponential growth
of the nonlocal density correlation is then apparent. Moreover, the
spatial pattern is determined by $\sigma_{a}(x)$, i.e., by a combination
of the wave-functions of the dominant mode.

A conclusion particularly relevant to the comparison with the non-stationary
configuration must be singled out. Namely, in the stationary regime,
the existence of instabilities does not lead the quantum fluctuations
to contribute to the mean density. In contrast, a self-amplifying
effect of quantum noise is apparent in the nonlocal density correlation
function.

\section*{Acknowledgments}

One of us (JMGL) acknowledges the support of the Spanish Ministerio
de Economía y Competitividad and the European Regional Development
Fund (Grant No. FIS 2016-79596-P).

\end{document}